\documentclass[%
reprint,
 amsmath,amssymb,
aps,
]{revtex4-2}

\usepackage{graphicx} 
\usepackage{dcolumn }
\usepackage{bm} 
\usepackage[hidelinks]{hyperref} 

\begin{document}


\title{Randium: A minimal model of universal viscous liquid dynamics}

\author{Ulf R. Pedersen}
\email{ulf@urp.dk}
\affiliation{%
 ``Glass and Time'', IMFUFA, Department of Science and Environment, Roskilde University, P.O. Box 260, DK-4000 Roskilde, Denmark
}%

\date{\today}

\begin{abstract}
When liquids are cooled and crystallization is avoided, their dynamics slow dramatically and the material eventually solidifies into an amorphous glass. Experiments show that chemically distinct glass forming liquids share universal features in both the spectral shape and the temperature dependence of the primary structural relaxation. We introduce Randium, a generic, energetically coarse-grained minimal model of viscous liquids. The model, inspired by results from atomistic molecular-dynamics simulations, is implemented on a two-dimensional lattice with Gaussian-distributed nearest-neighbor interactions. Temperature is the only control parameter, and at low temperatures, dynamic facilitation and dynamical heterogeneity emerge from simple nearest-neighbor rearrangements. The relaxation spectra obey time-temperature superposition, and they reproduce shapes observed experimentally for chemically distinct systems. The temperature dependence of the structural-relaxation time follows parabolic scaling, and the relaxation time grows exponentially with the heterogeneity length scale.
The mean-squared displacement collapses onto the universal master curve of the random barrier model. The absence of elasticity-induced facilitation in Randium shows that this is not required for universal viscous-liquid dynamics. Other explanations for universal relaxation are discussed in light of Randium.
\end{abstract}



\maketitle

Molecular motion becomes slower as liquids are cooled. If crystallization is bypassed, the system solidifies into a disordered structure called a glass
\cite{Dyre2007}. At the glass transition temperature, the
viscosity becomes so large that the liquid ceases to flow.
Various experiments have suggested that chemically distinct glass-forming liquids exhibit generic dynamics in their relaxation spectra and temperature dependence of structural relaxation. Depolarized dynamic light-scattering experiments by Pabst et al.  \cite{Pabst2021, Bhomer2025} provide striking evidence for a longstanding hypothesis: the spectral shape associated with structural relaxation in molecular liquids can be collapsed onto a common shape \cite{Kohlrausch1854, Cole1941, Davidson1950, Williams1970, Jonscher1977, Dyre1988, Dyre2000, Bierwirth2017, Zhang2017, Niss2018, Pabst2021, Ritort2003, Denny2003, Bhomer2025, Ong2024, Dyre2024, Lam2025, Railton2025}. 

In this paper, we aim to provide a framework for explaining the generic relaxation of highly viscous liquids.
%
%
%
To this end, we are inspired by atomic and molecular simulations \cite{Schroder2000, Scalliet2022, Heuer2008, Sciortino2005}, which show that, when the glass transition is approached, particles (atoms, molecules, or colloids) become temporarily confined on short timescales in a ``cage'' formed by their neighbors and can only rattle within it.
On a longer timescale, a small group of neighboring particles may move in a coordinated way, temporarily shifting their positions relative to one another. These collective flow events can allow particles to escape their cage, however, most rearrangements are reversible: after a brief displacement the particles return to essentially the same local configuration and no lasting transport occurs. In rare cases, a rearrangement does not reverse. A sequence of such mutually facilitating events can build up into a cascade that carries the system over a free-energy barrier, at which point the structure changes enough that the system loses memory of its initial configuration.

Due to the separation of timescales between cage vibrations and collective rearrangements, the dynamics can be viewed as jumping between local minima in a coarse-grained energy landscape \cite{Goldstein1969}. In this picture, the dynamics are described as a complex Markov chain of fundamental flow events.
%
We propose the following minimal criteria for a model of the energy landscape of a viscous liquid: 
i) The thermodynamics of the model should capture the inherent energies of real systems, typically Gaussian \cite{Sciortino2005, Heuer2008};
ii) The model should have a sense of space, capturing that fundamental flows are confined to a local rearrangement \cite{Goldstein1969,Donati1998,Vogel2004,Ediger2000,Tanaka2025,Dyre1987, Bouchaud1992, Dyre1995, Monthus1996};
iii) Dynamics should be an intrinsic property, i.e., independent of system size; and
iv) There should only be a single thermodynamic parameter in accordance with  isomorph theory \cite{Bailey2008, Gnan2009, Gundermann2011, Schroder2014}.
We conjecture that these rules constitute a family of models with universal viscous liquid dynamics.
Below, we construct such a model.

A central question is whether a simple model with the above features can capture the physics of real molecular systems. Competing views emphasize long-range elasticity-induced facilitation in glass-forming dynamics \cite{Dyre2006,Lematre2014,Phan2018,Ozawa2023,Scalliet2021,Hasyim2024,Costigliola2024}, a mechanism excluded from our minimal rules. Below we introduce Randium (an idealized model lacking \emph{long-ranged elastic facilitation}) and show that it reproduces generic dynamics of highly viscous glass-forming liquids. Thus, elasticity is not required to explain universal viscous liquid dynamics. In Randium, free-energy barriers (or traps), spatial dynamic facilitation, and dynamical heterogeneity emerge without being imposed. This differs from kinetically constrained models \cite{Ritort2003}, which enforce local kinetic constraints. It also differs from trap models \cite{Dyre1987, Bouchaud1992, Dyre1995, Monthus1996}, which assume an a priori distribution of trap depths. By reproducing glass-forming universality without explicit constraints, Randium provides an intrinsic, energetically grounded route to facilitation and heterogeneity.
The following sections define the model, present the results, and discuss their implications relative to other proposed mechanisms.
%

\section{Randium}

\begin{figure}
    \centering
    \includegraphics[width=0.6\linewidth]{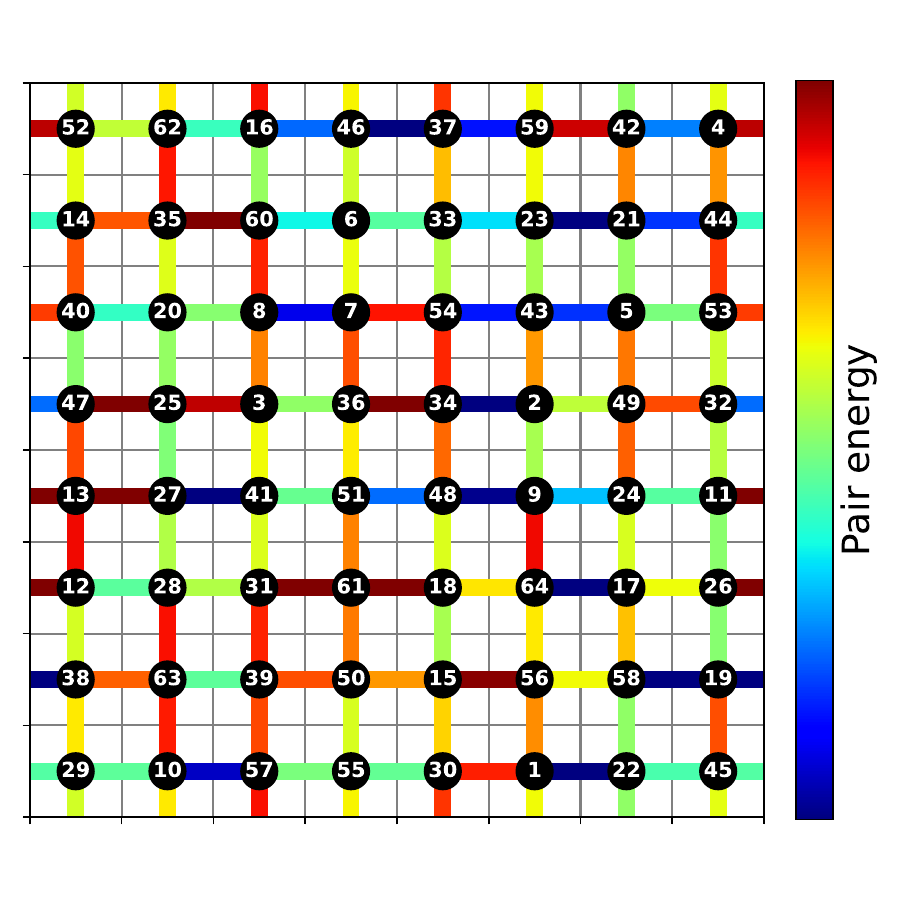}
    \caption{Illustration of the Randium model. The values inside each particle represent the particle type. The color of the line between neighbour particles represents the energy of that type-pair. For clarity, this figure shows an $8\times8$ lattice while the presented results are for a $192\times192$ lattice.}
    \label{fig:image_of_model}
\end{figure}

Consider a two-dimensional square lattice  with periodic boundary conditions (Fig.\ \ref{fig:image_of_model}). Let there be $L$ lattice points in each direction, and populate each point with one particle, so the total number of particles is $N=L^2$.
Let $(x_n, y_n)$ be the position of particle $n$.
Assign it the type $m_n$ out of a total number of $M$ types giving $\Omega = N!/((N/M)!)^M$ microstates.
The energy of a microstate is given by the sum of $2N$ interactions between nearest neighbors in the lattice.
%
%
On this level, each lattice site represents a local inherent structure of the underlying molecular, atomic, or colloidal system. A neighbor interaction encodes the interaction between two local inherent structures, i.e., the free-energy cost associated with their mutual arrangement. This provides a coarse-grained representation of the energy landscape in which the total energy is a sum of local contributions. Because each local configuration reflects many microscopic degrees of freedom, the resulting effective interactions are taken to be random and Gaussian distributed in agreement with observations in molecular simulations \cite{Sciortino2005, Heuer2008}. In short, the complex energy landscape is replaced by a spatially organized network of random energies \cite{Derrida1980, Derrida1981} --- replacing complexity with randomness.
To this aim, we define an $M\times M$ interaction matrix $I$ where elements are drawn from the standard normal distribution,
\begin{equation}\label{eq:standard_normal distribution}
P(I_{uv})=\exp(-I_{uv}^2/2)/\sqrt{2\pi},
\end{equation}
while ensuring that the interaction matrix is symmetric $I_{uv}=I_{vu}$. Without loss of generality, we use natural units where the standard deviation of the energy distribution is one.
The Hamiltonian can then be written as
\begin{equation}
H = \sum_{\langle ij \rangle} I_{m_im_j}.
\end{equation}
where $m_i$ is the type of the particle at position $(x,y)$ and $m_j$ is the type of one of the four nearest neighbors. In the limit where both $N$ and $M$ go to infinity, Randium exhibits trivial Gaussian thermodynamics \cite{Derrida1980, Derrida1981}, with an expected energy given by $\langle E \rangle = -2N\beta$ where $\beta$ is the inverse temperature. We note that for real systems the Gaussian is an approximation with a possible cutoff at low energies \cite{Saksaengwijit2004} that may result in an ideal glass state \cite{Kauzmann1948}.

Dynamics is defined through Monte Carlo (MC) simulations with nearest-neighbor swap attempts, employing Boltzmann's acceptance criterion. 
A nearest-neighbor particle swap represents a local collective rearrangement connecting two inherent states of the underlying fine-grained system.
(In a fine-grained reference
system, such a rearrangement typically involves tens of particles  \cite{Vogel2004}). This results in local rearrangements where back-jumps are likely. The unit of time is defined as one attempt per particle of the model.

Conveniently, the system can be efficiently equilibrated when $N=M$ with \emph{unphysical} swaps of particle identities (i.e. including MC attempts where particle type is changed). Both types of dynamics can be implemented using a parallelizable algorithm, allowing for efficient calculations on a graphics processing unit. This is essential, since it allows for the study of long timescales canonical for viscous liquid dynamics.
Below we present results with local particle swaps for a system size of $N=M=36\,864$ ($L=192$) using between two and 512 independent initial configurations. For this system size, one million swap attempts per particle on an NVIDIA GeForce RTX 4070 take about 5 minutes. A Python implementation is available at the DOI \href{https://doi.org/10.5281/zenodo.17554510}{10.5281/zenodo.17554510}.

Before continuing our investigation of the properties of Randium, we note that the framework can be generalized to other spatial dimensions, along with corresponding rules for connecting neighboring states. We leave such investigations to future studies. 



\begin{figure}
    \centering
    \includegraphics[width=0.99\linewidth]{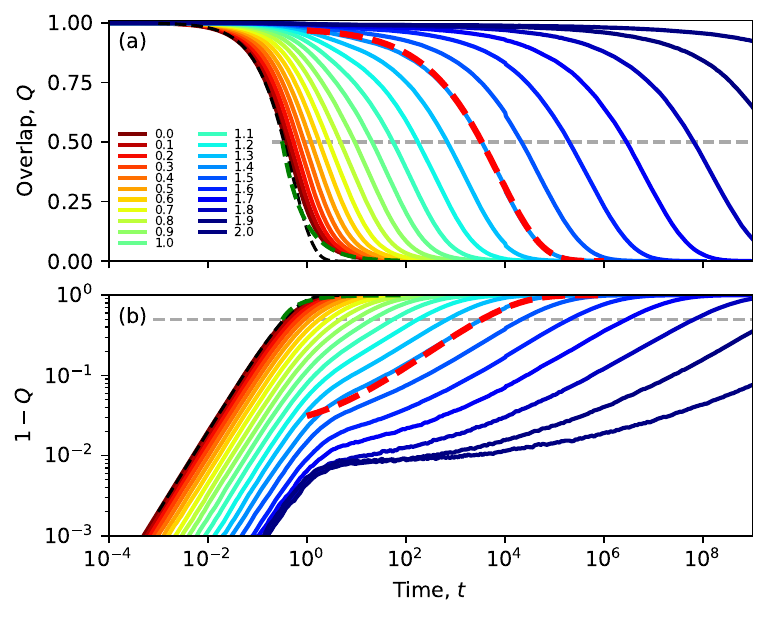}
    \caption{(a) Overlap order-parameter, $Q(t)$, as a function of time for inverse temperatures ranging from $\beta=0.0$ (dark red) to $\beta=2.0$ (dark blue). A characteristic relaxation time, $\tau$, is defined as where the overlap order-parameter is $\frac{1}{2}$ (gray dashed). At high temperatures (reddish colors), the relaxation is near exponential (black dashed): $\exp(-2t)$ (see Appendix \ref{sec:highT} for high-temperature predictions). At low temperatures (bluish colors), the relaxation is closer to a stretched exponential with exponent $\frac{1}{2}$ (red dashed): $A\exp(-\sqrt{t/t_0})$ where $A=0.98$, $t_0=\tau/[\ln(2A)]^2$ and $Q(\tau)=\frac{1}{2}$ (red dashed: $t_0=7600$ matching $Q(t)=0.5$ for $\beta=1.4$).
    (b) $\log(1-Q(t))$. 
    }
    \label{fig:overlap_order_parameter}
\end{figure}

\begin{figure}
    \centering
\includegraphics[width=0.99\linewidth]{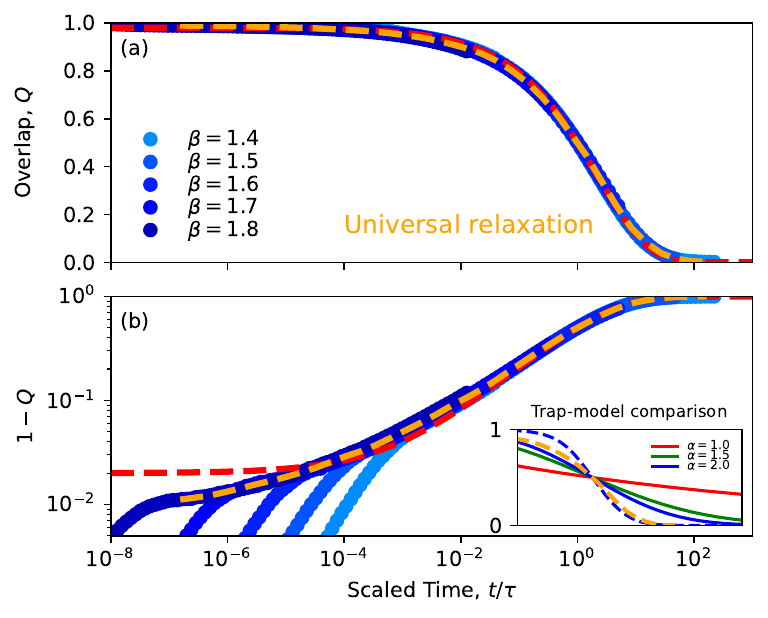}
    \caption{(a) The overlap order-parameter $Q(t/\tau)$ and (b) $\log(1-Q(t/\tau))$ as a function of scaled time. The orange dashed curve indicates a universal curve that $Q(t)$ approaches at intermediate and long times. For comparison, the red dashed is a stretch exponential (same as red dashed on Fig.\ \ref{fig:overlap_order_parameter}). The universal relaxation in not a stretch exponential (but it serves as an useful proxy). The inset compares Randium (dashed orange) to the mean persistence of the trap model (TM)  \cite{Dyre1987, Dyre1995,Bouchaud1992,Monthus1996} with the energy ($E>0$) distribution $P(E)=\exp(-E^\alpha)$ (solid red: $\alpha=1.0$, $\beta=0.955$, $\tau=1.4\times10^{6}$; solid green: $\alpha=1.5$, $\beta=3.2$, $\tau=1.7\times10^{6}$; solid blue: $\alpha=2.0$, $\beta=5.5$, $\tau=1.2\times10^{6}$; dashed blue: $\alpha=2.0$, $\beta=2.0$, $\tau=5.4$). The relaxation of TM differs from that of Randium (for the investigated parameters), since TM exhibit fat-tailed dynamics at low temperatures due to the absence of facilitation, which inhibit relaxation of low-energy traps \cite{Scalliet2021,Hasyim2024,Costigliola2024}.}
    \label{fig:overlap_order_scaled}
\end{figure}

\begin{figure}
    \centering
\includegraphics[width=0.99\linewidth]{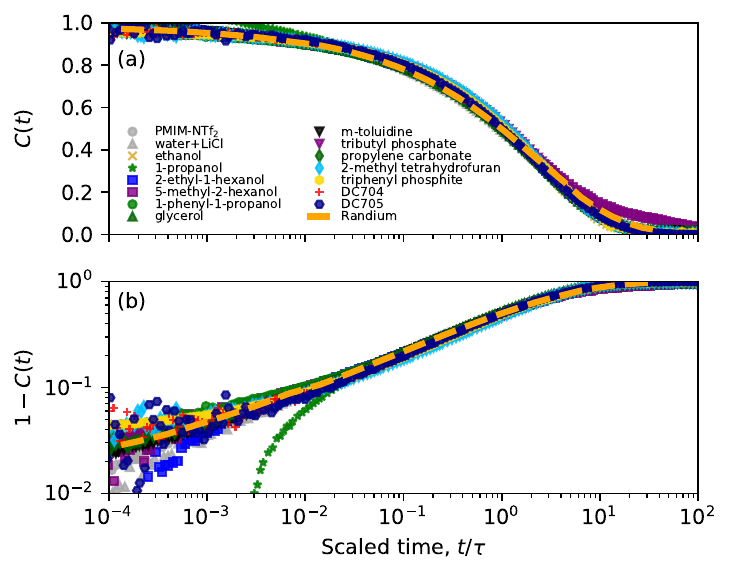}
    \caption{Comparing the relaxation of Randium (orange dashed) with molecules measured by depolarized dynamic light scattering \cite{Pabst2021, Bhomer2025}. 
    For the experimental data, $C(t)$ is the macroscopic dipole correlation, while for Randium $C(t)=Q(t)$. In both cases it quantifies relaxation, decreasing from 1 to 0 as memory is lost. The agreement is excellent.}
    \label{fig:comparisons}
\end{figure}

\begin{figure}
    \centering
\includegraphics[width=0.99\linewidth]{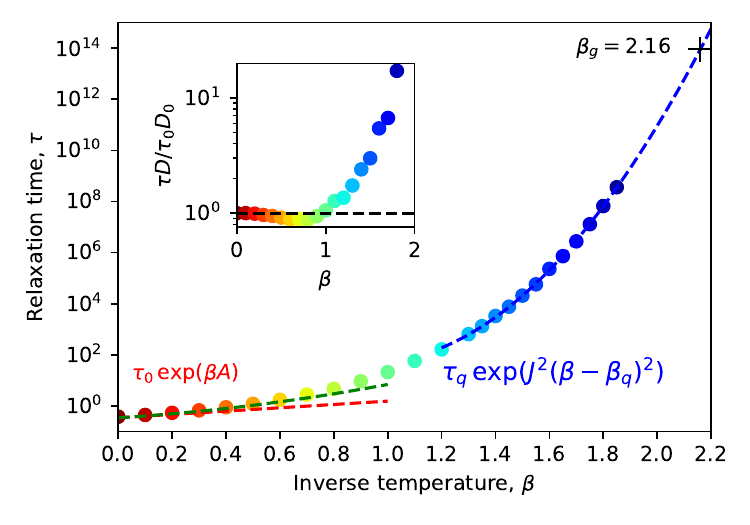}
    \caption{Temperature dependence of the relaxation time, $\tau=\tau(\beta)$. The red ($\tau_0=\ln(2)$, $A=3/2$) and green dashed lines are predictions for the high temperature limit, see Appendix \ref{sec:highT}. The blue dashed line is a parabolic scaling of kinetically constrained models \cite{Elmatad2009}, $\tau_q\exp(J^2[\beta-\beta_q]^2)$, with
    $\tau_q=50(1)$, $J=4.3(1)$, $\beta_q=0.93(3)$ (parentheses indicate the error on the final digit). By extrapolating, the inverse glass-transition temperature is estimated to $\beta_g=2.16$ (defined as $\tau(\beta_g)=10^{14}$). The inset shows decoupling of two timescales, here half-life $\tau$ and self-diffusion $D$, at low temperatures ($\beta>1$).}
    \label{fig:relaxation_times}
\end{figure}

\begin{figure}
    \centering
    \includegraphics[width=0.9\linewidth]{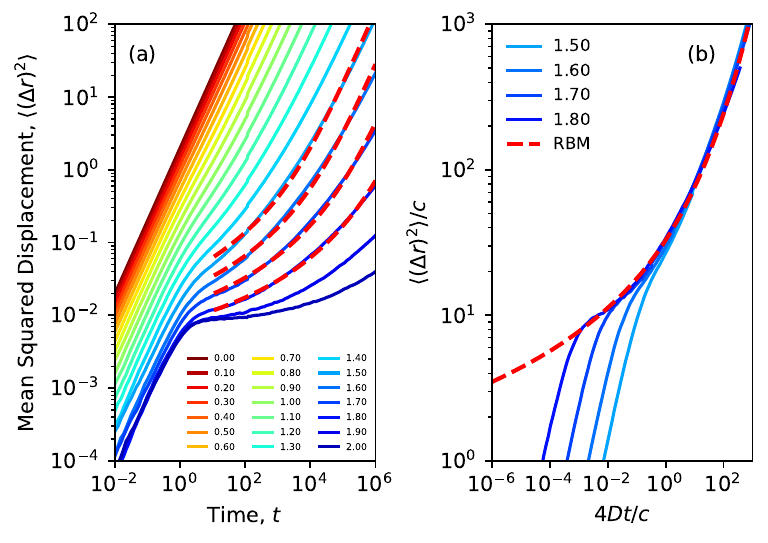}
    \caption{(a) Mean-squared displacement of Randium particles for $\beta = 0.0$ (dark red) to $\beta = 2.0$ (dark blue). The dashed red lines show the analytical MSD of the random barrier model (RBM) (see Ref.~\cite{Schrder2008}). (b) RBM scaling plot of the Randium MSD, demonstrating excellent agreement between Randium and the RBM. The axes are scaled using the diffusion coefficient $D$ and a scaling parameter $c$. The diffusion coefficient is obtained from the long-time limit of the MSD, while $c$ is chosen to provide the best overall collapse of the data.}
    \label{fig:msd}
\end{figure}

\begin{figure*}
    \centering
    \includegraphics[width=0.9\linewidth]{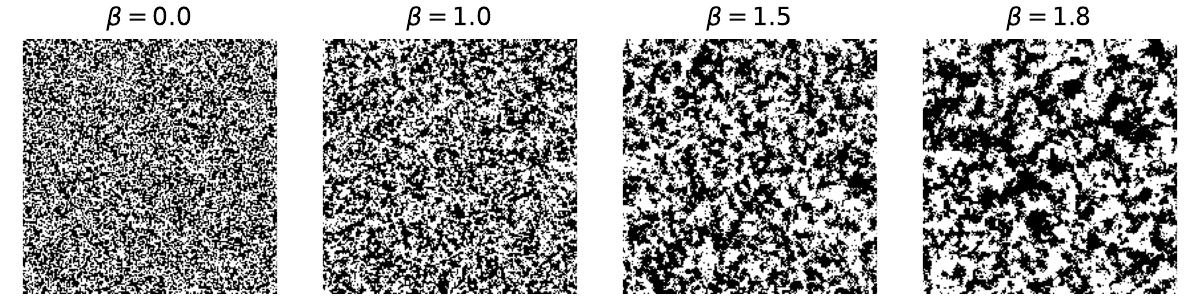}
    \caption{Spatial distribution of relaxed regions at $t=\tau$ for a range of $\beta$ values. Black corresponds to a site where the particle type has changed, and white to a site where it is unchanged.}
    \label{fig:overlap_half}
\end{figure*}

\begin{figure}
    \centering
    \includegraphics[width=0.9\linewidth]{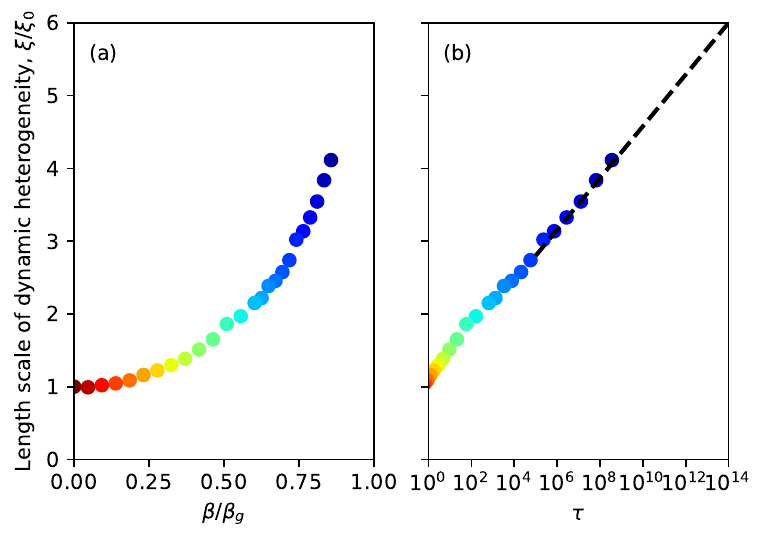}
    \caption{(a) Reduced length scale $\xi/\xi_0$ ($\xi_0=\xi(\beta=0)$) of dynamical heterogeneity at $t=\tau$ (Fig. \ref{fig:overlap_half}) estimated
from an exponential fit ($\exp(-r/\xi)$) to the ``spin-spin'' correlation function, $G(r)=\langle\sigma_{i,j}\sigma_{i,j+r}\rangle$ where $\sigma_{i,j}=-1$ if the site is unchanged, and $\sigma_{i,j}=+1$ otherwise (inspired by analysis of the 2D Ising model \cite{Shrock2022}). $\xi/\xi_0$ is shown versus reduced inverse temperature $\beta/\beta_g$. (b) The reduced length scale ($\xi/\xi_0$) as a function of relaxation time $\log(\tau)$. The black dashed line is a $\tau\propto\exp(\xi)$ fit, suggesting that that $\xi/\xi_0=6$ at the glass-transition temperature ($\tau_g=10^{14}$).}
    \label{fig:length_scales}
\end{figure}

\section{Results}

To monitor dynamics, we define the overlap, $O(t)$, as the fraction of sites that are occupied by the same particle after a time interval $t$. Let $Q(t)=\langle O(t)\rangle$ be the overlap function averaged over initial configurations. Fig.\ \ref{fig:overlap_order_parameter}(a) shows $Q(t)$ for inverse temperatures ($\beta$'s) ranging from zero (dark red) to two (dark blue). At high temperatures, the relaxation is nearly exponential (black dashed), whereas at low temperatures, it resembles a stretched exponential with an exponent of $\frac{1}{2}$ (red dashed): $A \exp(-\sqrt{t/t_0})$, where $A \simeq 0.98$ and $t_0$ is highly dependent on $\beta$. In the frequency domain, this corresponds to a minimum slope of the main relaxation peak of $-\frac{1}{2}$, consistent with dielectric experiments \cite{Nielsen2009} and found in one-dimensional kinetic constraint models \cite{Bramson1988}.
At the lowest temperatures, the empirical stretched-exponential fit suggest that a plateau develops due to particle back-jumps. This explains why $A$ is slightly less than one. Interestingly, as shown in Fig.\ \ref{fig:overlap_order_parameter}(b) by plotting $\log(1-Q(t))$ there is no universal plateau value.

In agreement with experimental results \cite{Niss2018}, the dynamics at low temperatures shows time-temperature superposition.
To show this, we define a characteristic relaxation time $\tau$ (referred to as half-life in the following) as the time, when half of the lattice sites (on average) have undergone a change, 
\begin{equation}\label{eq:tau}
   Q(\tau)=\frac{1}{2}.
\end{equation}
Fig.\ \ref{fig:overlap_order_scaled}(a) shows that for the lowest investigated temperatures, $Q(t/\tau)$ collapses to a universal relaxation curve (orange dashed).
Figure \ref{fig:overlap_order_scaled}(b) displays $[1-Q(t / \tau)]$ on a logarithmic scale. Interestingly, the scale-invariant relaxation curve is not a stretched exponential (compare to the red dashed curve).

How does the shape of the relaxation curve of Randium compare to the generic relaxation of experimental data on molecules? To answer this, we reanalyze depolarized dynamic light scattering data presented in Ref.\ \cite{Pabst2021, Bhomer2025}. Figs.\ \ref{fig:comparisons}(a) and \ref{fig:comparisons}(b) show that the empirical data follow the universal curve of Randium. Elmatad, Chandler, and Garrahan \cite{Elmatad2009} have shown that, at low temperatures, the relaxation-time of molecular systems follows a parabolic scaling, $\tau(T)=\tau_0\exp(J^2(\beta-\beta_0)^2)$ in agreement with Randium, see blue dashed line on Fig.\ \ref{fig:relaxation_times}. The dynamical range from high-temperature dynamics to the glass-transition for molecular liquids typically spans $15$ orders of magnitude ($10^{-13}$ s at high temperature, to $10^{2}$ s at the glass-transition temperature). From this we estimate the inverse glass-transition temperature of Randium to $\beta_g=2.16$ (see $+$ on Fig.\ \ref{fig:relaxation_times}). The Angell fragility index at the glass-transition temperature, here defined as $m\equiv \left.\frac{d\log_{10} \tau}{d[\beta/\beta_g]}\right|_{\beta_g}$ giving $m=2J^2\beta_g^2(1-\beta_q/\beta_g)/\ln 10$, is 43 -- within the range of typical molecular glass-formers (this value will likely depend on lattice connectivity).

Figure \ref{fig:msd}(a) shows the mean-squared displacement (MSD) for a range of temperatures. At low temperatures, the MSD displays the characteristic caging behavior of glass-forming systems. Surprisingly, the dynamics are accurately captured by the random barrier model (RBM), despite the absence of explicit barriers in the model construction. Figure \ref{fig:msd}(b) demonstrates this by showing that the RBM scaling procedure collapses all MSD data onto a common master curve.

In summary, Randium reproduces time-temperature superposition, the universal relaxation spectrum and the universal shape of the structural relaxation time ($\tau(\beta)$) of molecular systems. We refer to these properties as the \emph{intrinsic viscous liquid dynamics}.

\section{Discussion}
Why does a simple model, here exemplified with Randium, reproduce the intrinsic viscous liquid dynamics of molecular systems? To answer this, recall that dynamical heterogeneity \cite{SchmidtRohr1991, Ediger2000, Tanaka2025} plays a crucial role in understanding viscous liquid dynamics. Specifically, at low temperatures, there are regions of space where particles relax quickly, and regions where structural changes are more sluggish. This results in dynamical heterogeneity-induced decoupling of timescale (exemplified by Stokes-Einstein breakdown \cite{Fujara1992}), at low temperatures, as reproduced by Randium: The inset in Fig.\ \ref{fig:relaxation_times} illustrates this. The panels in Fig.\ \ref{fig:overlap_half} show the sites whose particle type has changed (black) after a time $t=\tau$. Interestingly, as temperature is lowered (increase of $\beta$), the cooperatively rearranging regions increase in size, as suggested by Adam and Gibbs \cite{Adam1965}. Figure \ref{fig:length_scales}(a) shows that the characteristic length-scale $\xi$ increase more than a factor of four in the investigated temperature range (see the legend of Fig.~\ref{fig:length_scales} for the definition of $\xi$). To a good approximation, the relaxation time scales as $\tau\propto\exp(\xi/\xi_0)$. An extrapolation suggests that at the glass-transition temperature, the length-scale is increased by a factor of six.

What is the origin of dynamic heterogeneity in Randium? 
To answer this question, consider a low-temperature configuration with favorable nearest-neighbor interactions. When two particles swap, each of them retains one neighbor but gains three new ones, whose interaction energies are drawn from the  $P(I_{uv})$ distribution and are therefore likely to be unfavorable at the given temperature. Thus, particles tend to swap back -- the particles are \emph{trapped}  \cite{Dyre1987, Bouchaud1992, Dyre1995, Monthus1996}. However, in rare events, the initial swap may facilitate nearby swaps, allowing particles to find new neighbours with favorable pair energies (escaping the trap). This will involve the rearrangement of a region of particles, creating an region of enhanced mobility. This mobile region may facilitate dynamics in nearby regions since particles in that region now have new possibilities of meeting new neighbours with potentially favorable neighbours.
This is similar to what happens in a molecular liquid, where local rearrangements of molecules can trigger cascades of cooperative motion, leading to regions of high mobility embedded in an otherwise rigid structure.

How does Randium compare to other proposed explanations of generic viscous liquid dynamics?
Historically, the first descriptions are empirical approaches such as fits to a stretched exponential \cite{Kohlrausch1854, Williams1970} in the time-domain, or the Cole-Cole fit in the frequency-domain \cite{Cole1941, Davidson1950}.
More theoretically founded approaches include
kinetic constraint models (KMC) \cite{Lam2025},
spin-glass models \cite{Edwards1975, Sherrington1975, Derrida1980, Hopfield1982, Parisi1979, Nishikawa2020, Sherrington2025, Dahlberg2025}, energetic barrier and trap models \cite{Dyre1988, Bouchaud1992, Monthus1996, Diezemann1997, Monthus2003, Schrder2008, Scalliet2021}, elastic models \cite{Dyre2006, Vasin2022, Hasyim2024}, and the recently proposed Hyper-sphere model \cite{Railton2025}. 
Randium builds on the idea of an intrinsic energy landscape put forward by Goldstein in 1969 \cite{Goldstein1969}. 
In particular, the distinguishable-particle lattice (DPL) model by Lam and coworkers \cite{Zhang2017, Ong2024, Lee2025}, and the lattice-gas on a random energy landscape in three dimensions \cite{Pedersen2006} are closely related. Like Randium, these models are defined as particles on a lattice -- unlike Randium, dynamics are defined as particles moving into a void, like in the kids' toy \emph{15-Puzzle} \cite{Ratner1990}. The motivation for this dynamics is string-like motions seen in computational studies of atomic glass-formers \cite{Donati1998, Schroder2000}. Randium is unpretentious in the sense that it eliminates both the explicit kinetic constraints of KMC models and the void defects inherent to the DPL model. Randium is characterized by a single control parameter (temperature) which makes it an ideal candidate a minimal-model \cite{Niss2018} within isomorph theory \cite{Gnan2009, Gundermann2011, Schroder2014} and single-parameter aging \cite{Hecksher2015, Hecksher2024}. We leave such investigations to future work.



In summary, we have introduced Randium -- a minimal model grounded in physical insights from atomistic simulations and experiments. We have shown that Randium successfully reproduces the intrinsic viscous liquid dynamics observed in such systems. This suggests that Randium belongs to a broader class of models governed by similar physics. We conjecture that this class includes variations of Randium with different connectivity, such as a simple cubic lattice in three dimensions, as well as alternative distributions of neighbour interaction energies. We conjecture that the precise realization of a Randium-like system has little influence on the universal dynamical behavior, aside from trivial scaling factors.
Finally, since Randium is significantly simpler than the inherent energy landscape that can be constructed molecular systems, it offers the possibility of connecting to more fundamental, analytically tractable models \cite{Parisi1979, Bouchaud1992, Schrder2008, Scalliet2021}. In this sense, Randium serves as a stepping-stone framework, bridging realistic molecular models with highly idealized approaches such as the random barrier model (see Fig.\ \ref{fig:msd}), the trap model (see insert on Fig. \ref{fig:overlap_order_scaled}(b)), or kinetically constrained models (blue dashed on Fig.~\ref{fig:relaxation_times}).
We leave such investigations to future studies.

\section*{Data and code availability}
Data and a Python implementation of Randium are available from Zenodo at DOI \href{https://doi.org/10.5281/zenodo.17554510}{10.5281/zenodo.17554510}. The repository contains additional figures showing the determination of fit-parameters in Fig.\ \ref{fig:relaxation_times}, and fits to determine the $\xi$ length-scales presented in Fig. \ref{fig:length_scales}.

\begin{acknowledgments}
The author thanks Thomas Blochowicz, Mark Ediger, Tina Hecksher, Andreas Heuer, Camille Scalliet, Walter Kob, Peter Harrowell, Nicholas Bailey, Thomas B. Schrøder, and Jeppe C. Dyre for helpful discussions, and in particular Florian Pabst for curating and discussing experimental data.
\end{acknowledgments}

\appendix

\section{High temperature dynamics}
\label{sec:highT}
To make a theoretical prediction for the half-life $\tau$ at infinite temperature ($\beta=0$), we may consider the dynamics of Randium as a lattice gas of non-interacting particles. In each step, two sites out of $N\gg 1$ are selected at random and swapped. Thus, the probability that a given site participates in an update is $\frac{2}{N}$, while the probability that it remains untouched is $1-\frac{2}{N}$. After $k$ steps, the probability that a given site has not yet been updated is
\begin{equation}\label{Eq:Qk}
Q(k) = \left(1-\frac{2}{N}\right)^k.
\end{equation}
For $N\to\infty$ this simplifies to
\begin{equation}\label{eq:Qt_theory}
    Q(t) = \exp(-2t) \quad (t\ll 1)
\end{equation}
where $t\equiv k/N$ is the definition of time (see black dashed line on Fig.\ \ref{fig:overlap_order_parameter}). At long times a given particle makes a random walk on a square lattice, and $Q(t)$ is given by the return probability of a two-dimensional Gauss-distribution,
\begin{equation}\label{eq:Qt_theory_longtime}
    Q(t) = \frac{1}{2\pi t} \quad (t\gg 1).
\end{equation}
Let $\tau_0$ be the time required for half of the sites to remain unvisited (at $\beta=0$), i.e. $Q(\tau_0) = \frac{1}{2}$. From Eq.\ (\ref{eq:Qt_theory}) we get
\begin{equation}
\tau_0 = \ln(\sqrt{2}) \simeq 0.35 \quad (\beta=0) .
\end{equation}
To provide a description for the inverse temperature ($\beta$) dependence we assume an Arrhenius law,
$\tau = \tau_0\exp(\beta A)$. We find  empirically that $A=3/2$,
\begin{equation}
\tau = \ln(\sqrt{2})\exp\left(3\beta/2\right)\quad(\beta\to0),
\end{equation}
see red dashed line on Fig.\ \ref{fig:relaxation_times}.
A more accurate empirical description is
\begin{equation}
\tau = \ln(\sqrt{2})\exp\left(3(\beta+\beta^2)/2\right)\quad(\beta\to0).
\end{equation}
See green dashed line on Fig.\ \ref{fig:relaxation_times}.

\bibliography{references}

@article{Dyre1987,
  title = {Master-equation appoach to the glass transition},
  volume = {58},
  ISSN = {0031-9007},
  DOI = {10.1103/physrevlett.58.792},
  number = {8},
  journal = {Phys. Rev. Lett.},
  author = {Dyre,  Jeppe C.},
  year = {1987},
  month = Feb,
  pages = {792–795}
}

@article{Dyre1995,
  title = {Energy master equation: A low-temperature approximation to B\"{a}ssler’s random-walk model},
  volume = {51},
  ISSN = {1095-3795},
  url = {http://dx.doi.org/10.1103/PhysRevB.51.12276},
  DOI = {10.1103/physrevb.51.12276},
  number = {18},
  journal = {Phys. Rev. B},
  publisher = {American Physical Society (APS)},
  author = {Dyre,  Jeppe C.},
  year = {1995},
  month = May,
  pages = {12276–12294}
}

@article{Phan2018,
  title = {Elastically Collective Nonlinear Langevin Equation Theory of Glass-Forming Liquids: {T}ransient Localization,  Thermodynamic Mapping,  and Cooperativity},
  volume = {122},
  ISSN = {1520-5207},
  DOI = {10.1021/acs.jpcb.8b04975},
  number = {35},
  journal = {J. Phys. Chem. B},
  publisher = {American Chemical Society (ACS)},
  author = {Phan,  Anh D. and Schweizer,  Kenneth S.},
  year = {2018},
  month = Aug,
  pages = {8451–8461}
}

@article{Gnan2009,
  title = {Pressure-energy correlations in liquids. IV. “Isomorphs” in liquid phase diagrams},
  volume = {131},
  ISSN = {1089-7690},
  DOI = {10.1063/1.3265957},
  number = {23},
  journal = {J. Chem. Phys.},
  publisher = {AIP Publishing},
  author = {Gnan,  Nicoletta and Schrøder,  Thomas B. and Pedersen,  Ulf R. and Bailey,  Nicholas P. and Dyre,  Jeppe C.},
  year = {2009},
  month = dec 
}

@article{Schroder2014,
  title = {Simplicity of condensed matter at its core: Generic definition of a Roskilde-simple system},
  volume = {141},
  ISSN = {1089-7690},
  DOI = {10.1063/1.4901215},
  number = {20},
  journal = {J. Chem. Phys.},
  publisher = {AIP Publishing},
  author = {Schrøder,  Thomas B. and Dyre,  Jeppe C.},
  year = {2014},
  month = nov 
}

@article{Bailey2008,
doi = {10.1088/0953-8984/20/24/244113},
year = {2008},
month = {may},
volume = {20},
number = {24},
pages = {244113},
author = {Bailey, Nicholas P and Christensen, Tage and Jakobsen, Bo and Niss, Kristine and Boye Olsen, Niels and Pedersen, Ulf R and Schrøder, Thomas B and Dyre, Jeppe C},
title = {Glass-forming liquids: one or more ‘order’ parameters?},
journal = {J. Phys. Condens. Matter.},
}

@article{Hecksher2024,
  title = {Single parameter aging and density scaling},
  volume = {161},
  pages = 194504,
  ISSN = {1089-7690},
  DOI = {10.1063/5.0234620},
  number = {19},
  journal = {J. Chem. Phys.},
  publisher = {AIP Publishing},
  author = {Hecksher,  Tina and Niss,  Kristine},
  year = {2024},
  month = nov 
}

@article{Hecksher2015,
  title = {Communication: Direct tests of single-parameter aging},
  volume = {142},
  pages = 241103,
  ISSN = {1089-7690},
  DOI = {10.1063/1.4923000},
  number = {24},
  journal = {J. Chem. Phys.},
  publisher = {AIP Publishing},
  author = {Hecksher,  Tina and Olsen,  Niels Boye and Dyre,  Jeppe C.},
  year = {2015},
  month = jun 
}

@article{Gundermann2011,
  title = {Predicting the density-scaling exponent of a glass-forming liquid from Prigogine–Defay ratio measurements},
  volume = {7},
  ISSN = {1745-2481},
  DOI = {10.1038/nphys2031},
  number = {10},
  journal = {Nat. Phys.},
  publisher = {Springer Science and Business Media LLC},
  author = {Gundermann,  Ditte and Pedersen,  Ulf R. and Hecksher,  Tina and Bailey,  Nicholas P. and Jakobsen,  Bo and Christensen,  Tage and Olsen,  Niels B. and Schrøder,  Thomas B. and Fragiadakis,  Daniel and Casalini,  Riccardo and Michael Roland,  C. and Dyre,  Jeppe C. and Niss,  Kristine},
  year = {2011},
  month = jul,
  pages = {816–821}
}

@article{Lematre2014,
  title = {Structural Relaxation is a Scale-Free Process},
  volume = {113},
  ISSN = {1079-7114},
  DOI = {10.1103/physrevlett.113.245702},
  number = {24},
  journal = {Phys. Rev. Lett.},
  publisher = {American Physical Society (APS)},
  author = {Lemaître,  Anaël},
  year = {2014},
  month = dec 
}

@article{Scalliet2021,
  title = {Excess wings and asymmetric relaxation spectra in a facilitated trap model},
  volume = {155},
  pages = {064505},
  ISSN = {1089-7690},
  DOI = {10.1063/5.0060408},
  number = {6},
  journal = {J. Chem. Phys.},
  publisher = {AIP Publishing},
  author = {Scalliet,  Camille and Guiselin,  Benjamin and Berthier,  Ludovic},
  year = {2021},
  month = aug 
}

@article{Ozawa2023,
  title = {Elasticity,  Facilitation,  and Dynamic Heterogeneity in Glass-Forming Liquids},
  volume = {130},
  ISSN = {1079-7114},
  DOI = {10.1103/physrevlett.130.138201},
  number = {13},
  journal = {Phys. Rev. Lett.},
  publisher = {American Physical Society (APS)},
  author = {Ozawa,  Misaki and Biroli,  Giulio},
  year = {2023},
  month = mar 
}

@article{Costigliola2024,
  title = {Glass-forming liquids need facilitation},
  volume = {121},
  ISSN = {1091-6490},
  DOI = {10.1073/pnas.2408798121},
  number = {25},
  journal = {PNAS},
  publisher = {PNAS},
  author = {Costigliola,  Lorenzo and Hecksher,  Tina and Dyre,  Jeppe C.},
  year = {2024},
  month = jun 
}

@article{Ritort2003,
  title = {Glassy dynamics of kinetically constrained models},
  volume = {52},
  ISSN = {1460-6976},
  DOI = {10.1080/0001873031000093582},
  number = {4},
  journal = {Adv. Phys.},
  publisher = {Informa UK Limited},
  author = {Ritort,  F. and Sollich,  P.},
  year = {2003},
  month = jun,
  pages = {219–342}
}

@article{Bramson1988,
  title = {Asymptotic Behavior of Densities in Diffusion-Dominated Annihilation Reactions},
  volume = {61},
  ISSN = {0031-9007},
  DOI = {10.1103/physrevlett.61.2397},
  number = {21},
  journal = {Phys. Rev. Lett.},
  publisher = {American Physical Society (APS)},
  author = {Bramson,  Maury and Lebowitz,  Joel L.},
  year = {1988},
  month = nov,
  pages = {2397–2400}
}

@article{Nishikawa2020,
  title = {Lattice Glass Model in Three Spatial Dimensions},
  volume = {125},
  pages = 065501,
  ISSN = {1079-7114},
  DOI = {10.1103/physrevlett.125.065501},
  number = {6},
  journal = {Phys. Rev. Lett.},
  publisher = {American Physical Society (APS)},
  author = {Nishikawa,  Yoshihiko and Hukushima,  Koji},
  year = {2020},
  month = aug 
}

@article{Dyre2007,
  title = {Ten themes of viscous liquid dynamics},
  volume = {19},
  ISSN = {1361-648X},
  DOI = {10.1088/0953-8984/19/20/205105},
  number = {20},
  journal = {J. Phys. Condens. Matter},
  publisher = {IOP Publishing},
  author = {Dyre,  Jeppe C},
  year = {2007},
  month = apr,
  pages = {205105}
}

@article{Niss2018,
  title = {Perspective: Searching for simplicity rather than universality in glass-forming liquids},
  volume = {149},
  ISSN = {1089-7690},
  DOI = {10.1063/1.5048093},
  number = {23},
  journal = {J. Chem. Phys.},
  publisher = {AIP Publishing},
  author = {Niss,  Kristine and Hecksher,  Tina},
  year = {2018},
  month = dec 
}

@article{Kauzmann1948,
author = {Kauzmann, Walter.},
title = {The Nature of the Glassy State and the Behavior of Liquids at Low Temperatures.},
journal = {Chem. Rev.},
volume = {43},
number = {2},
pages = {219-256},
year = {1948},
doi = {10.1021/cr60135a002},
}

@article{Adam1965,
  title = {On the Temperature Dependence of Cooperative Relaxation Properties in Glass-Forming Liquids},
  volume = {43},
  ISSN = {1089-7690},
  DOI = {10.1063/1.1696442},
  number = {1},
  journal = {J. Chem. Phys.},
  publisher = {AIP Publishing},
  author = {Adam,  Gerold and Gibbs,  Julian H.},
  year = {1965},
  month = jul,
  pages = {139–146}
}

@article{SchmidtRohr1991,
  title = {Nature of nonexponential loss of correlation above the glass transition investigated by multidimensional {NMR}},
  volume = {66},
  ISSN = {0031-9007},
  DOI = {10.1103/physrevlett.66.3020},
  number = {23},
  journal = {Phys. Rev. Lett.},
  publisher = {American Physical Society (APS)},
  author = {Schmidt-Rohr,  K. and Spiess,  H. W.},
  year = {1991},
  month = jun,
  pages = {3020–3023}
}

@article{Ediger2000,
  title = {Spatially Heterogeneous Dynamics in Supercooled Liquids},
  volume = {51},
  ISSN = {1545-1593},
  DOI = {10.1146/annurev.physchem.51.1.99},
  number = {1},
  journal = {Annu. Rev. Phys. Chem.},
  publisher = {Annual Reviews},
  author = {Ediger,  M. D.},
  year = {2000},
  month = oct,
  pages = {99–128}
}

@article{Tanaka2025,
  title = {Structural Origin of Dynamic Heterogeneity in Supercooled Liquids},
  volume = {129},
  ISSN = {1520-5207},
  DOI = {10.1021/acs.jpcb.4c06392},
  number = {3},
  journal = {J. Phys. Chem. B},
  publisher = {American Chemical Society (ACS)},
  author = {Tanaka,  Hajime},
  year = {2025},
  month = jan,
  pages = {789–813}
}

@article{Fujara1992,
  title = {Translational and rotational diffusion in supercooled orthoterphenyl close to the glass transition},
  volume = {88},
  ISSN = {1434-6036},
  DOI = {10.1007/bf01323572},
  number = {2},
  journal = {Z. Phys.},
  publisher = {Springer Science and Business Media LLC},
  author = {Fujara,  F. and Geil,  B. and Sillescu,  H. and Fleischer,  G.},
  year = {1992},
  month = jun,
  pages = {195–204}
}

@article{Zhang2017,
  title = {Emergent facilitation behavior in a distinguishable-particle lattice model of glass},
  volume = {95},
  ISSN = {2469-9969},
  url = {http://dx.doi.org/10.1103/PhysRevB.95.184202},
  DOI = {10.1103/physrevb.95.184202},
  number = {18},
  journal = {Phys. Rev. B},
  publisher = {American Physical Society (APS)},
  author = {Zhang,  Ling-Han and Lam,  Chi-Hang},
  year = {2017},
  month = may 
}

@article{Ong2024,
  title = {Relating fragile-to-strong transition to fragile glass via lattice model simulations},
  volume = {109},
  ISSN = {2470-0053},
  DOI = {10.1103/physreve.109.054124},
  number = {5},
  journal = {Phys. Rev. E},
  publisher = {American Physical Society (APS)},
  author = {Ong,  Chin-Yuan and Lee,  Chun-Shing and Gao,  Xin-Yuan and Zhai,  Qiang and Yu,  Zhenhao and Shi,  Rui and Deng,  Hai-Yao and Lam,  Chi-Hang},
  year = {2024},
  month = may 
}

@article{Lee2025,
  title = {Unified picture of structural relaxation,  beta relaxation, and excess wing in glass formers},
  volume = {112},
  pages = 045422,
  ISSN = {2470-0053},
  DOI = {10.1103/kx3x-c8n1},
  number = {4},
  journal = {Phys. Rev. E},
  publisher = {American Physical Society (APS)},
  author = {Lee,  Chun-Shing and Deng,  Hai-Yao and Zhang,  Linghan and Xiao,  Chu and Li,  Bo and Yip,  Cho-Tung and Lam,  Chi-Hang},
  year = {2025},
  month = oct 
}

@article{Lam2025,
  title = {Emergent facilitation by random constraints in a facilitated random walk model of glass},
  volume = {111},
  ISSN = {2470-0053},
  url = {http://dx.doi.org/10.1103/PhysRevE.111.044120},
  DOI = {10.1103/physreve.111.044120},
  number = {4},
  journal = {Phys. Rev. E},
  publisher = {American Physical Society (APS)},
  author = {Lam,  Leo S. I. and Deng,  Hai-Yao and Zhang,  Wei-Bing and Nwankwo,  Udoka and Xiao,  Chu and Yip,  Cho-Tung and Lee,  Chun-Shing and Ruan,  Haihui and Lam,  Chi-Hang},
  year = {2025},
  month = apr 
}

@article{Hasyim2024,
  title = {Emergent facilitation and glassy dynamics in supercooled liquids},
  volume = {121},
  ISSN = {1091-6490},
  url = {http://dx.doi.org/10.1073/pnas.2322592121},
  DOI = {10.1073/pnas.2322592121},
  number = {23},
  journal = {PNAS},
  publisher = {Proceedings of the National Academy of Sciences},
  author = {Hasyim,  Muhammad R. and Mandadapu,  Kranthi K.},
  year = {2024},
  month = may 
}

@article{Elmatad2009,
  title = {Corresponding States of Structural Glass Formers},
  volume = {113},
  ISSN = {1520-5207},
  DOI = {10.1021/jp810362g},
  number = {16},
  journal = {J. Phys. Chem. B},
  publisher = {American Chemical Society (ACS)},
  author = {Elmatad,  Yael S. and Chandler,  David and Garrahan,  Juan P.},
  year = {2009},
  month = mar,
  pages = {5563–5567}
}

@article{Bhomer2025,
  title = {On the spectral shape of the structural relaxation in supercooled liquids},
  volume = {162},
  ISSN = {1089-7690},
  DOI = {10.1063/5.0254534},
  number = {12},
  journal = {J. Chem. Phys.},
  publisher = {AIP Publishing},
  author = {B\"{o}hmer,  Till and Pabst,  Florian and Gabriel,  Jan Philipp and Zeißler,  Rolf and Blochowicz,  Thomas},
  year = {2025},
  month = mar 
}

@article{Pabst2021,
  title = {Generic Structural Relaxation in Supercooled Liquids},
  volume = {12},
  ISSN = {1948-7185},
  DOI = {10.1021/acs.jpclett.1c00753},
  number = {14},
  journal = {J. Phys. Chem. Lett.},
  publisher = {American Chemical Society (ACS)},
  author = {Pabst,  Florian and Gabriel,  Jan Philipp and B\"{o}hmer,  Till and Weigl,  Peter and Helbling,  Andreas and Richter,  Timo and Zourchang,  Parvaneh and Walther,  Thomas and Blochowicz,  Thomas},
  year = {2021},
  month = apr,
  pages = {3685–3690}
}

@article{Dyre2024,
  title = {Solid-that-Flows Picture of Glass-Forming Liquids},
  volume = {15},
  ISSN = {1948-7185},
  DOI = {10.1021/acs.jpclett.3c03308},
  number = {6},
  journal = {J. Phys. Chem. Lett.},
  publisher = {American Chemical Society (ACS)},
  author = {Dyre,  Jeppe C.},
  year = {2024},
  month = feb,
  pages = {1603–1617}
}

@article{Bierwirth2017,
  title = {Generic Primary Mechanical Response of Viscous Liquids},
  volume = {119},
  ISSN = {1079-7114},
  DOI = {10.1103/physrevlett.119.248001},
  number = {24},
  journal = {Phys. Rev. Lett.},
  publisher = {American Physical Society (APS)},
  author = {Bierwirth,  S. Peter and B\"{o}hmer,  Roland and Gainaru,  Catalin},
  year = {2017},
  month = dec 
}

@article{Nielsen2009,
  title = {Prevalence of approximate $\sqrt{t}$ relaxation for the dielectric $\alpha$ process in viscous organic liquids},
  volume = {130},
  ISSN = {1089-7690},
  DOI = {10.1063/1.3098911},
  number = {15},
  journal = {J. Chem. Phys.},
  publisher = {AIP Publishing},
  author = {Nielsen,  Albena I. and Christensen,  Tage and Jakobsen,  Bo and Niss,  Kristine and Olsen,  Niels Boye and Richert,  Ranko and Dyre,  Jeppe C.},
  year = {2009},
  month = apr 
}

@article{Kohlrausch1854,
  author    = {Rudolf Kohlrausch},
  title     = {Theorie des elektrischen Rückstandes in der Leidner Flasche},
  journal   = {Pogg. Ann.},
  volume    = {91},
  pages     = {56--82},
  year      = {1854},
  doi       = {10.1002/andp.18541670203}
}

@article{Cole1941,
  author    = {Kenneth S. Cole and Robert H. Cole},
  title     = {Dispersion and Absorption in Dielectrics {I. A}lternating Current Characteristics},
  journal   = {J. Chem. Phys.},
  volume    = {9},
  number    = {4},
  pages     = {341--351},
  year      = {1941},
  doi       = {10.1063/1.1750906}
}

@article{Davidson1950,
  author    = {D. W. Davidson and R. H. Cole},
  title     = {Dielectric Relaxation in Glycerine},
  journal   = {J. Chem. Phys.},
  volume    = {18},
  number    = {10},
  pages     = {1417--1419},
  year      = {1950},
  doi       = {10.1063/1.1747496}
}

@article{Williams1970,
  author    = {Graham Williams and David C. Watts},
  title     = {Non-symmetrical dielectric relaxation behaviour arising from a simple empirical decay function},
  journal   = {Trans. Faraday Soc.},
  volume    = {66},
  pages     = {80--85},
  year      = {1970},
  doi       = {10.1039/TF9706600080}
}

@article{Jonscher1977,
  title = {The ‘universal’ dielectric response},
  volume = {267},
  ISSN = {1476-4687},
  DOI = {10.1038/267673a0},
  number = {5613},
  journal = {Nature},
  publisher = {Springer Science and Business Media LLC},
  author = {Jonscher,  A. K.},
  year = {1977},
  month = jun,
  pages = {673–679}
}

@article{Dyre2000,
  title = {Universality of ac conduction in disordered solids},
  volume = {72},
  ISSN = {1539-0756},
  DOI = {10.1103/revmodphys.72.873},
  number = {3},
  journal = {Rev. Mod. Phys.},
  publisher = {American Physical Society (APS)},
  author = {Dyre,  Jeppe C. and Schrøder,  Thomas B.},
  year = {2000},
  month = jul,
  pages = {873–892}
}

@article{Denny2003,
  author  = {Denny, R. Aldrin and Reichman, David R. and Bouchaud, Jean-Philippe},
  title   = {Trap Models and Slow Dynamics in Supercooled Liquids},
  journal = {Phys. Rev. Lett.},
  volume  = {90},
  pages   = {025503},
  year    = {2003},
  doi     = {10.1103/PhysRevLett.90.025503}
}

@article{Pedersen2006,
  title = {An energy landscape model for glass-forming liquids in three dimensions},
  volume = {352},
  ISSN = {0022-3093},
  url = {http://dx.doi.org/10.1016/j.jnoncrysol.2006.03.127},
  DOI = {10.1016/j.jnoncrysol.2006.03.127},
  number = {42–49},
  journal = {J. Non-Cryst. Solids},
  publisher = {Elsevier BV},
  author = {Pedersen,  Ulf R. and Hecksher,  Tina and Dyre,  Jeppe C. and Schrøder,  Thomas B.},
  year = {2006},
  month = nov,
  pages = {5210–5215}
}

@article{Goldstein1969,
  title = {Viscous Liquids and the Glass Transition{: A} Potential Energy Barrier Picture},
  volume = {51},
  ISSN = {1089-7690},
  DOI = {10.1063/1.1672587},
  number = {9},
  journal = {J. Chem. Phys.},
  publisher = {AIP Publishing},
  author = {Goldstein,  Martin},
  year = {1969},
  month = nov,
  pages = {3728–3739}
}

@article{Schroder2000,
  title = {Crossover to potential energy landscape dominated dynamics in a model glass-forming liquid},
  volume = {112},
  ISSN = {1089-7690},
  DOI = {10.1063/1.481621},
  number = {22},
  journal = {J. Chem. Phys.},
  publisher = {AIP Publishing},
  author = {Schrøder,  Thomas B. and Sastry,  Srikanth and Dyre,  Jeppe C. and Glotzer,  Sharon C.},
  year = {2000},
  month = jun,
  pages = {9834–9840}
}

@article{Donati1998,
  title = {Stringlike Cooperative Motion in a Supercooled Liquid},
  volume = {80},
  ISSN = {1079-7114},
  DOI = {10.1103/physrevlett.80.2338},
  number = {11},
  journal = {Phys. Rev. Lett.},
  publisher = {American Physical Society (APS)},
  author = {Donati,  Claudio and Douglas,  Jack F. and Kob,  Walter and Plimpton,  Steven J. and Poole,  Peter H. and Glotzer,  Sharon C.},
  year = {1998},
  month = mar,
  pages = {2338–2341}
}

@article{Sciortino2005,
  title = {Potential energy landscape description of supercooled liquids and glasses},
  volume = {2005},
  ISSN = {1742-5468},
  DOI = {10.1088/1742-5468/2005/05/p05015},
  journal = {J. Stat. Mech.},
  publisher = {IOP Publishing},
  author = {Sciortino,  Francesco},
  year = {2005},
  month = may,
  pages = {P05015}
}

@article{Heuer2008,
  title = {Exploring the potential energy landscape of glass-forming systems: from inherent structures via metabasins to macroscopic transport},
  volume = {20},
  ISSN = {1361-648X},
  DOI = {10.1088/0953-8984/20/37/373101},
  number = {37},
  journal = {J. Phys. Condens. Matter},
  publisher = {IOP Publishing},
  author = {Heuer,  Andreas},
  year = {2008},
  month = aug,
  pages = {373101}
}

@article{Saksaengwijit2004,
  title = {Origin of the Fragile-to-Strong Crossover in Liquid Silica as Expressed by its Potential-Energy Landscape},
  volume = {93},
  ISSN = {1079-7114},
  DOI = {10.1103/physrevlett.93.235701},
  number = {23},
  journal = {Phys. Rev. Lett.},
  publisher = {American Physical Society (APS)},
  author = {Saksaengwijit,  A. and Reinisch,  J. and Heuer,  A.},
  year = {2004},
  month = dec 
}

@article{Shrock2022,
  title = {On general-n coefficients in series expansions for row spin–spin correlation functions in the two-dimensional Ising model},
  volume = {55},
  ISSN = {1751-8121},
  DOI = {10.1088/1751-8121/ac9654},
  number = {42},
  journal = {J. Phys. A},
  publisher = {IOP Publishing},
  author = {Shrock,  Robert},
  year = {2022},
  month = oct,
  pages = {425001}
}

@article{Derrida1980,
  title = {Random-Energy Model{: Limit} of a Family of Disordered Models},
  volume = {45},
  ISSN = {0031-9007},
  doi = {10.1103/physrevlett.45.79},
  number = {2},
  journal = {Phys. Rev. Lett},
  publisher = {American Physical Society (APS)},
  author = {Derrida,  B.},
  year = {1980},
  month = jul,
  pages = {79–82}
}

@article{Derrida1981,
  title = {Random-energy model{: An} exactly solvable model of disordered systems},
  volume = {24},
  ISSN = {0163-1829},
  DOI = {10.1103/physrevb.24.2613},
  number = {5},
  journal = {Phys. Rev. B},
  publisher = {American Physical Society (APS)},
  author = {Derrida,  B.},
  year = {1981},
  pages = {2613–2626}
}

@article{Edwards1975,
  title = {Theory of spin glasses},
  volume = {5},
  ISSN = {0305-4608},
  DOI = {10.1088/0305-4608/5/5/017},
  number = {5},
  journal = {J. Phys. F: Met. Phys.},
  publisher = {IOP Publishing},
  author = {Edwards,  S F and Anderson,  P W},
  year = {1975},
  month = may,
  pages = {965–974}
}

@article{Sherrington1975,
  title = {Solvable Model of a Spin-Glass},
  volume = {35},
  ISSN = {0031-9007},
  DOI = {10.1103/physrevlett.35.1792},
  number = {26},
  journal = {Phys. Rev. Lett.},
  publisher = {American Physical Society (APS)},
  author = {Sherrington,  David and Kirkpatrick,  Scott},
  year = {1975},
  month = dec,
  pages = {1792–1796}
}

@article{Parisi1979,
  title = {Infinite Number of Order Parameters for Spin-Glasses},
  volume = {43},
  ISSN = {0031-9007},
  DOI = {10.1103/physrevlett.43.1754},
  number = {23},
  journal = {Phys. Rev. Lett.},
  publisher = {American Physical Society (APS)},
  author = {Parisi,  G.},
  year = {1979},
  month = dec,
  pages = {1754–1756}
}

@article{Hopfield1982,
  title = {Neural networks and physical systems with emergent collective computational abilities},
  volume = {79},
  ISSN = {1091-6490},
  DOI = {10.1073/pnas.79.8.2554},
  number = {8},
  journal = {PNAS},
  publisher = {Proceedings of the National Academy of Sciences},
  author = {Hopfield,  J J},
  year = {1982},
  month = apr,
  pages = {2554–2558}
}

@article{Sherrington2025,
  title = {50 years of spin glass theory
},
  DOI = {10.48550/arXiv.2505.24432},
  journal = {preprint on arXiv.org; doi:},
  author = {David Sherrington and Scott Kirkpatrick},
  year = {2025},
}

@article{Dahlberg2025,
  title = {Spin-glass dynamics: Experiment, theory, and simulation},
  url = {https://arxiv.org/abs/2412.08381},
  journal = {Rev. Mod. Phys.},
  note = {Accepted 8 August, 2025. doi: 10.1103/ctp2-zwyr},
  author = {E.D. Dahlberg and I. González-Adalid Pemartín and E. Marinari and V. Martin-Mayor and J. Moreno-Gordo and R.L. Orbach and I. Paga and G. Parisi and F. Ricci-Tersenghi and J.J. Ruiz-Lorenzo and D. Yllanes},
}

@article{Dyre1988,
  title = {The random free-energy barrier model for ac conduction in disordered solids},
  volume = {64},
  ISSN = {1089-7550},
  url = {http://dx.doi.org/10.1063/1.341681},
  DOI = {10.1063/1.341681},
  number = {5},
  journal = {J. Appl. Phys.},
  publisher = {AIP Publishing},
  author = {Dyre,  Jeppe C.},
  year = {1988},
  month = sep,
  pages = {2456–2468}
}

@article{Schrder2008,
  title = {{ac H}opping Conduction at Extreme Disorder Takes Place on the Percolating Cluster},
  volume = {101},
  ISSN = {1079-7114},
  DOI = {10.1103/physrevlett.101.025901},
  number = {2},
  journal = {Phys. Rev. Lett.},
  publisher = {American Physical Society (APS)},
  author = {Schrøder,  Thomas B. and Dyre,  Jeppe C.},
  year = {2008},
  month = jul,
pages = {025901}
}

@article{Monthus1996,
  title = {Models of traps and glass phenomenology},
  volume = {29},
  ISSN = {1361-6447},
  DOI = {10.1088/0305-4470/29/14/012},
  number = {14},
  journal = {J. Phys. A},
  publisher = {IOP Publishing},
  author = {Monthus,  Cécile and Bouchaud,  Jean-Philippe},
  year = {1996},
  pages = {3847–3869}
}

@article{Diezemann1997,
  title = {A free-energy landscape model for primary relaxation in glass-forming liquids: Rotations and dynamic heterogeneities},
  volume = {107},
  ISSN = {1089-7690},
  DOI = {10.1063/1.474148},
  number = {23},
  journal = {J. Chem. Phys.},
  publisher = {AIP Publishing},
  author = {Diezemann,  Gregor},
  year = {1997},
  month = dec,
  pages = {10112–10120}
}

@article{Monthus2003,
  title = {Anomalous diffusion,  localization,  aging,  and subaging effects in trap models at very low temperature},
  volume = {68},
  ISSN = {1095-3787},
  DOI = {10.1103/physreve.68.036114},
  number = {3},
  journal = {Phys. Rev. E},
  publisher = {American Physical Society (APS)},
  author = {Monthus,  Cécile},
  year = {2003},
  month = sep,
pages = "036114"
}

@article{Bouchaud1992,
  title = {Weak ergodicity breaking and aging in disordered systems},
  volume = {2},
  ISSN = {1286-4862},
  DOI = {10.1051/jp1:1992238},
  number = {9},
  journal = {J. Phys. {I}},
  publisher = {EDP Sciences},
  author = {Bouchaud,  J. P.},
  year = {1992},
  month = sep,
  pages = {1705–1713}
}

@article{Railton2025,
  title = {Viscous liquid dynamics modeled as random walks within overlapping hyperspheres},
  volume = {111},
  ISSN = {2470-0053},
  DOI = {10.1103/physreve.111.055301},
  number = {5},
  journal = {Phys. Rev. E},
  publisher = {American Physical Society (APS)},
  author = {Railton,  Mark F. B. and Uhre,  Eva and Dyre,  Jeppe C. and Schrøder,  Thomas B.},
  year = {2025},
  month = may 
}

@article{Dyre2006,
  title = {Colloquium: The glass transition and elastic models of glass-forming liquids},
  volume = {78},
  ISSN = {1539-0756},
  DOI = {10.1103/revmodphys.78.953},
  number = {3},
  journal = {Rev. Mod. Phys.},
  publisher = {American Physical Society (APS)},
  author = {Dyre,  Jeppe C.},
  year = {2006},
  month = sep,
  pages = {953–972}
}

@article{Vasin2022,
  title = {Glass transition as a topological phase transition},
  volume = {106},
  ISSN = {2470-0053},
  url = {http://dx.doi.org/10.1103/PhysRevE.106.044124},
  DOI = {10.1103/physreve.106.044124},
  number = {4},
  journal = {Phys. Rev. E},
  publisher = {American Physical Society (APS)},
  author = {Vasin,  M. G.},
  year = {2022},
  month = oct,
pages = "044124"
}

@article{Scalliet2022,
  title = {Thirty Milliseconds in the Life of a Supercooled Liquid},
  volume = {12},
  ISSN = {2160-3308},
  DOI = {10.1103/physrevx.12.041028},
  number = {4},
  journal = {Phys. Rev. X},
  publisher = {American Physical Society (APS)},
  author = {Scalliet,  Camille and Guiselin,  Benjamin and Berthier,  Ludovic},
  year = {2022},
  month = dec,
pages="041028"
  
}

@article{Vogel2004,
  title = {Particle rearrangements during transitions between local minima of the potential energy landscape of a binary Lennard-Jones liquid},
  volume = {120},
  ISSN = {1089-7690},
  DOI = {10.1063/1.1644538},
  number = {9},
  journal = {J. Chem. Phys.},
  publisher = {AIP Publishing},
  author = {Vogel,  Michael and Doliwa,  Burkhard and Heuer,  Andreas and Glotzer,  Sharon C.},
  year = {2004},
  month = mar,
  pages = {4404–4414}
}

@article{Ratner1990,
  title = {The (n$^2$-1)-puzzle and related relocation problems},
  volume = {10},
  ISSN = {0747-7171},
  DOI = {10.1016/s0747-7171(08)80001-6},
  number = {2},
  journal = {J. Symb. Comput.},
  publisher = {Elsevier BV},
  author = {Ratner,  Daniel and Warmuth,  Manfred},
  year = {1990},
  month = aug,
  pages = {111-137}
}

\end{document}